# Modeling Hurricanes using Principle Component Analysis in conjunction with Non-Response Analysis

Rebecca D. Wooten, Joy D'Andrea

*Department of Mathematics and Statistics, University of South Florida, Tampa 33620*

**Abstract:** This paper demonstrates how principle component analysis can be used to determine the distinct factors that house the terms that explain the variance among the co-dependent variables and how non-response analysis can be applied to model the non-functional relationship that exist in a dynamic system. Moreover, the analysis indicates that there are pumping actions or ebb and flow between the pressure and the water temperature readings near the surface of the water days before a tropical storm forms in the Atlantic Basin and that there is a high correlation between storm conditions and buoy conditions three-four days before a storm forms. Further analysis shows that that the relationship among the variables is conical.

**Keywords:** Principle Component Analysis, Factor Analysis, Non-Response Analysis, Hurricanes, Modeling

## 1. Introduction

Principal Component Analysis (PCA) is used to reduce a large set of variables to a smaller set of variables with the same explanatory power. In this paper, PCA was used to reduce a large set of related terms recorded during storms (hurricanes) in the Atlantic Basin and buoy readings from 2000 – 2009 to determine the principle components. Once the distinct components are identified, non-response analysis can be used to model the non-functional relationship amongst the terms in the principle component. The data used in this analysis include measurements taken every three hours when a named storm is present.

The hurricane data used in this analysis are taken from UNISYS Weather Center from 2000-2009, Figure 1, and includes a time stamp, name of the hurricane, location (latitude and longitude) and the main variable of interest **wind speed** and **pressure**.

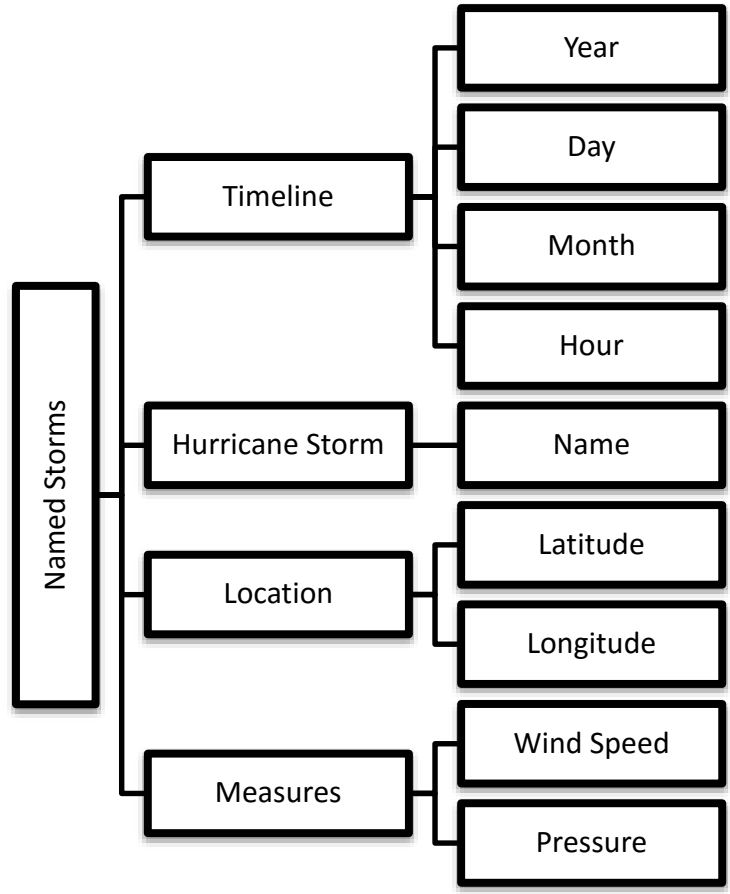

**Figure 1:** Data diagram of named storms in the Atlantic Basin

Based on the Wooten-Tsokos scale, there are many named storms that would not be considered true hurricanes since the pressures have not dropped to a point that hurricane force winds can be sustained, Figure 2. Among 4160 readings, there are 165 named storms: 16 storms classified at maximum wind speed as a tropical storm, 82 classified as category 1, 28 classified as category 2, 15 classified as category 3, 13 classified as category 4 and 10 classified as category 5. When considered by reading, nearly 30% are when the wind speeds are less than 45 knots (tropical). Only 19 readings show wind speeds greater than or equal to 145 knots.

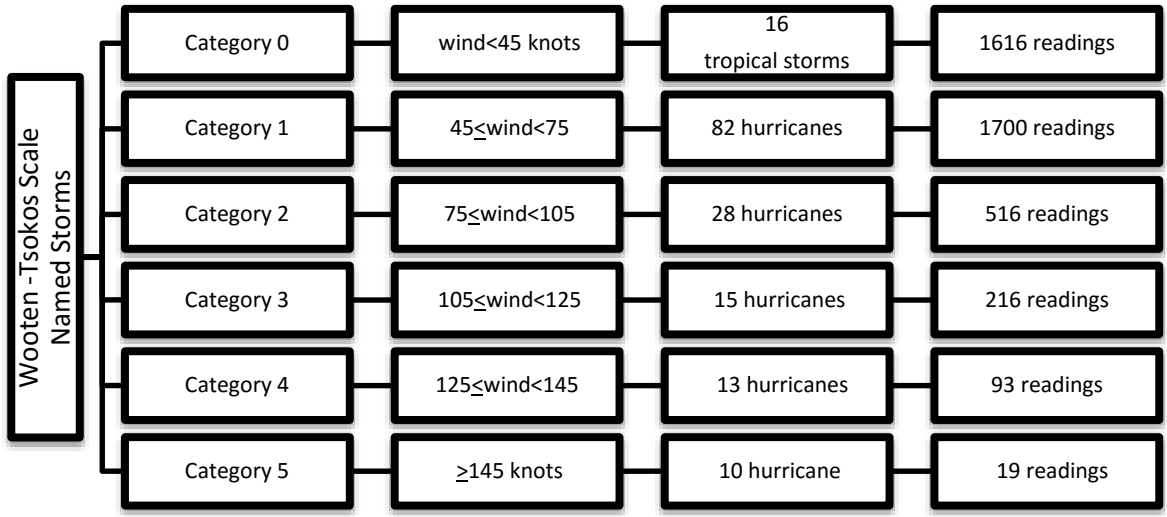

**Figure 2:** Hurricane classification based upon the Wooten-Tsokos Scale and breakdown of the data by classification and by number of readings in that category.

The readings include **wind speeds** and **pressures**, Figure 3, during the life cycle of a storm from tropical depression to hurricane status to dissipation. Among 4924 readings, the mean is 50 knot and a mode 30 knots.

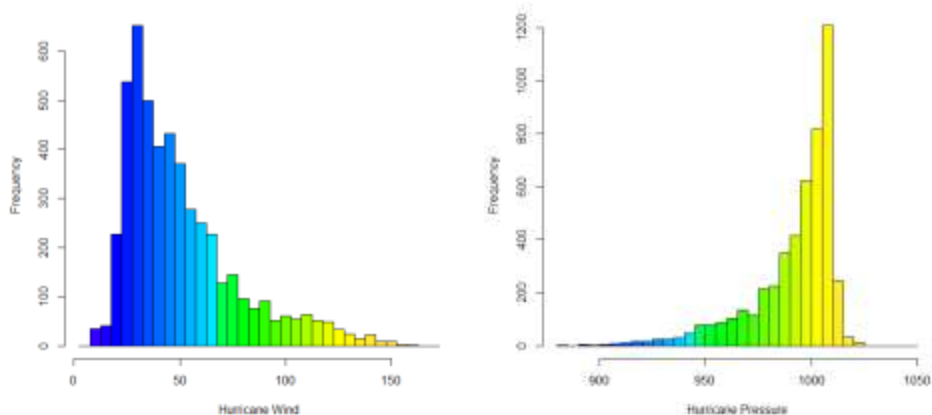

**Figure 3:** Histogram of **wind speed** and **pressure** as measured during a tropical storm.

A second data set, Figure 4, from the National Data Buoy Center containing the **wind speed**, **pressure**, **atmospheric temperature** and **water temperature** where added to the **wind speed** and **pressure** readings from the hurricanes with 36 daily time shifts used to measure the buoy conditions days before the formation of a tropical storm.

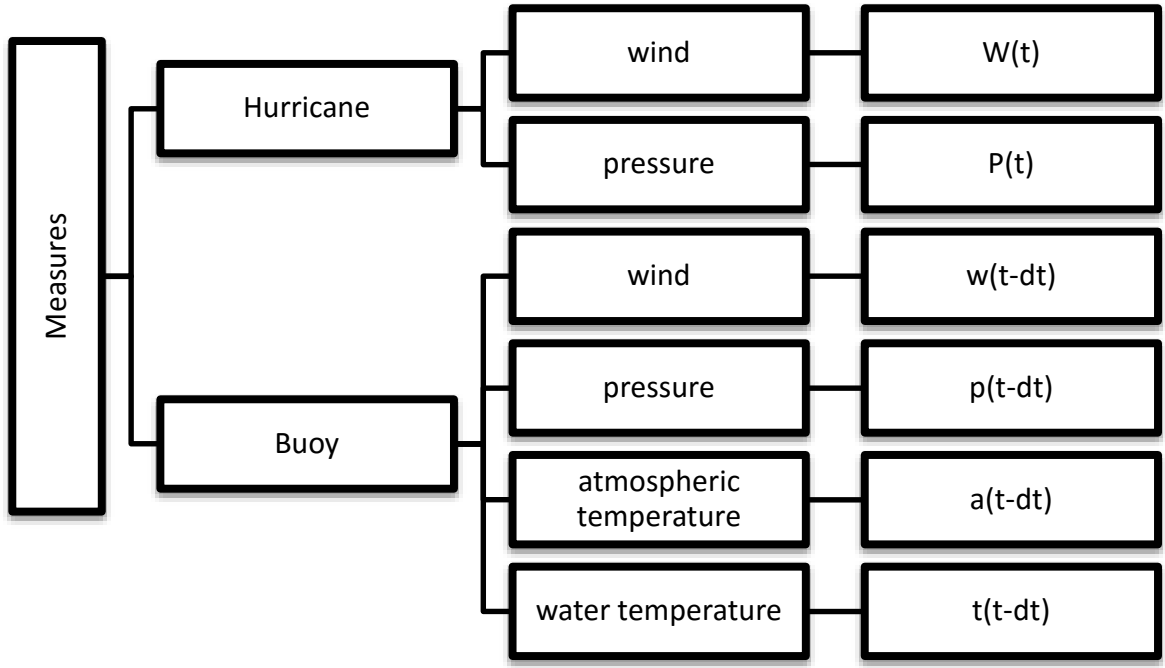

**Figure 4:**   Measured variables of interest including time shifts in the buoy conditions.

## 2. Principle Component Analysis

The terms to be considered using principle component analysis and non-response analysis includes the following 36 terms: the primary variables, the second degree terms and all first order interaction terms.

$$\{W, P, w, p, a, t, W^2, WP, Ww, Wp, P^2, \dots\}$$

The complete list is given in Table 1 which sorts the terms into factors. Using principle component analysis, we found four principle components. The fourth principle component indicates that pressure readings near the surface of the water at the buoy are extremely constant. When measuring the constant nature of a variable, the coefficient of determination approaches 1 as the variance approaches zero and is 0.75 for uniformly distributed measures. Measuring the constant nature of the variable, with coefficient of determinations of 0.9999852, 0.9876067, and 0.9919942, pressure, atmospheric temperature and water temperature, respectively, show little to no variability whereas the wind speed at the buoy has a coefficient of determination near 0.79; that is, with only 79% of the total sums of squares is explained by the mean is distribution of the wind speeds recorded at the buoy is nearly uniformly distributed.

**Table 1:** Loading Factors ordered by Factors and the percent variance contained in the individual components.

|       | Factor 1 | Factor 2 | Factor 3 | Factor 4 |
|-------|----------|----------|----------|----------|
| $W$   | 1        |          |          |          |
| $P$   | −0.94    |          |          |          |
| $W^2$ | 0.97     |          |          |          |
| $WP$  | 1        |          |          |          |
| $wW$  | 0.7      |          | 0.62     |          |
| $pW$  | 1        |          |          |          |
| $aW$  | 0.98     |          |          |          |
| $tW$  | 0.99     |          |          |          |
| $P^2$ | −0.94    |          |          |          |
| $pP$  | −0.92    |          |          |          |
| $a$   |          | 0.96     |          |          |
| $t$   |          | 0.96     |          |          |
| $aP$  |          | 0.96     |          |          |
| $tP$  |          | 0.95     |          |          |
| $ap$  |          | 0.96     |          |          |
| $tp$  |          | 0.97     |          |          |
| $a^2$ |          | 0.97     |          |          |
| $at$  |          | 0.98     |          |          |
| $t^2$ |          | 0.96     |          |          |
| $w$   |          |          | 0.97     |          |
| $wP$  |          |          | 0.97     |          |
| $w^2$ |          |          | 0.94     |          |
| $wp$  |          |          | 0.97     |          |
| $wa$  |          |          | 0.98     |          |
| $wt$  |          |          | 0.99     |          |
| $p$   |          | -0.3     |          | 0.91     |
| $p^2$ |          | -0.3     |          | 0.91     |

Table 2 gives the SS loading weights, the proportion of variance contained in each factor and the cumulative proportions; and indicates that four components (factors) were sufficient, explaining 96% of the variation. With a SS loading that is less than 1, the fifth factor was found to be insignificant. Whereas the first factor with an SS loading of 9.05 indicates that at least 34% of the variance among the terms exists.

**Table 2:** SS loading weights, proportion of variance explained by each of the factors and the cumulative proportion of explained variance.

|  | Factor1 | Factor2 | Factor3 | Factor4 | Factor5 |
|---|---|---|---|---|---|
| SS loadings | 9.05 | 8.77 | 6.35 | 1.84 | 0.37 |
| Proportion Variance | 0.34 | 0.32 | 0.24 | 0.07 | 0.01 |
| Cumulative Variance | 0.34 | 0.66 | 0.9 | 0.96 | 0.98 |

## 3. Non-Response Analysis

In this analysis, the terms of interest are those variables, interaction and second degree terms belonging to the first principle component and the primary variable of interest is **wind speed** of a hurricane as related to the pressure of the hurricane and the buoy conditions.

Consider the non-response model:

$$u = \alpha_1 W + \alpha_2 P + \alpha_3 W^2 + \alpha_4 P^2$$

$$+\alpha_5 Ww + \alpha_6 Wp + \alpha_7 Wa + \alpha_8 Wt + \alpha_9 WP + \alpha_{10} Pp$$

where $u$ for wind speed of hurricane as a function of the other measurements:

$$A = a_3, B = a_1 + a_5 w + a_6 p + a_7 a + a_8 t + \alpha_9 P,$$

$$C = a_2 P + a_4 P^2 + a_{10} Pp - unity$$

$$\widehat{W_L} = \frac{-B - \sqrt{B^2 - 4AC}}{2A}.$$

$$\widehat{W_U} = \frac{-B + \sqrt{B^2 - 4AC}}{2A}.$$

Solutions obtained are conditional to the pressure as this indirect relationship is co-dependent on volume by the Ideal Gas Law and Boltzmann's Equation.

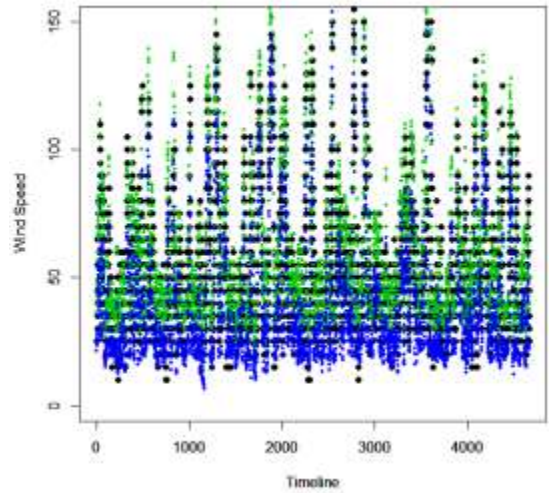

**Figure 5**: Line graph of **wind speed** recorded during a tropical storm/hurricane over time in hours past January 1, 2000.

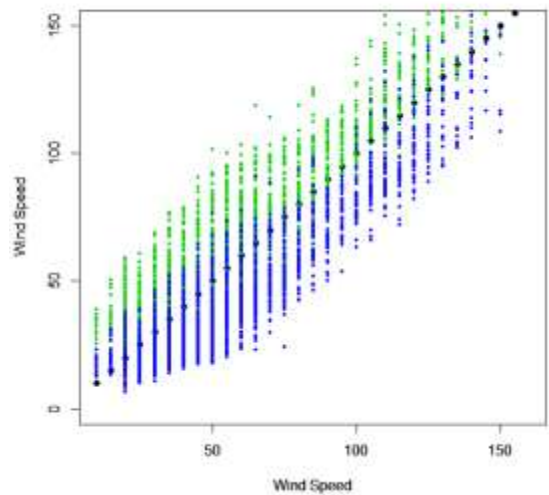

**Figure 6:** Scatterplot of **wind speed** and the limits for the estimated wind speed

To estimate the wind speed within a storm, $W$, let

$$\widehat{W} = \begin{cases} \widehat{W}_L & if \ |\widehat{W}_L - W| < |\widehat{W}_U - W| \\ \widehat{W}_U & otherwise \end{cases}$$

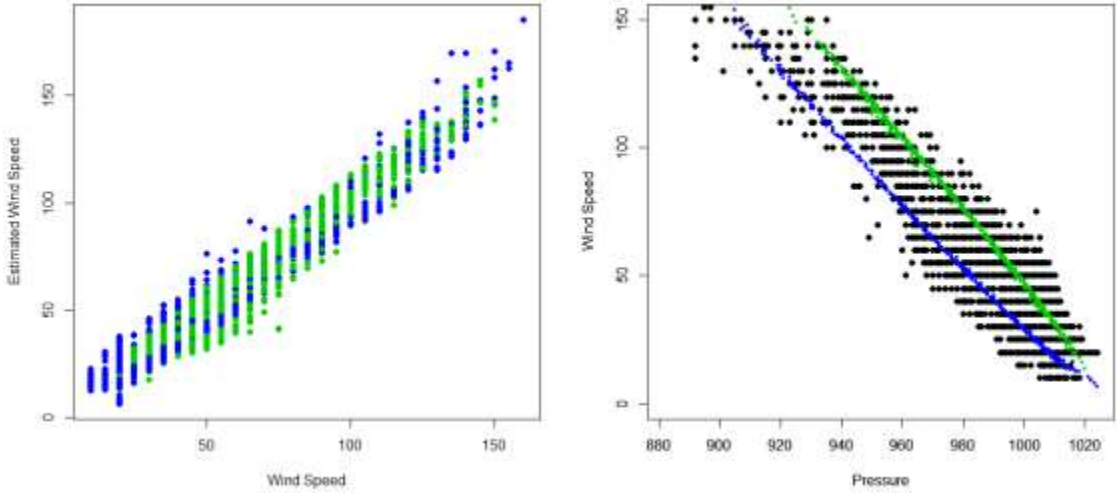

**Figure 7:** Scatterplot of (a) wind speed and the estimated wind speed and (b) wind speed and pressure both observed (black) and the upper and lower limits (green and blue).

The relationship between shifts as the storm hits its peak and starts to dissipate, Figure 8.

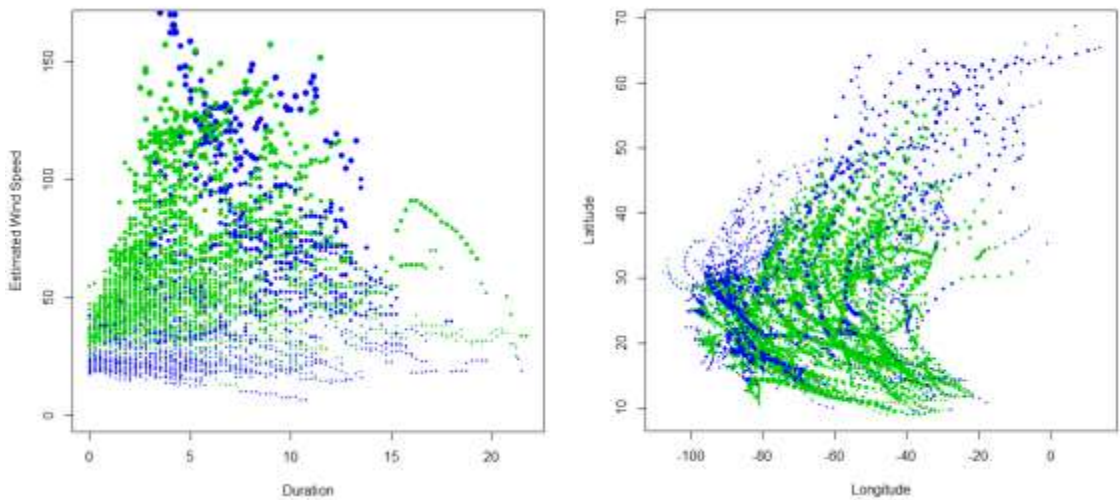

**Figure 8:** A time series of (a) estimated wind speed over time and (b) a scatter plot of latitude and longitude colored in terms of the upper and lower limits (green and blue).

To determine the number of days ($dt$) before the storms formation that best predicts the intensity of a storm by using the correlation between $W$ and $\widehat{W}$ is computed for

$$dt = 1,2,\ldots,36\ days.$$

A correlation was found to be 0.9882843 when $dt$ is three days; that is the buoy condition three days before the hurricane reading shows the highest correlation with the storm conditions. As illustrated in Figure 9, there is a sinusoidal relationship in the measured correlations; that is, there is a 'pumping action' or between wind speeds as measured in a storm and the conditions as measured near the surface of the water.

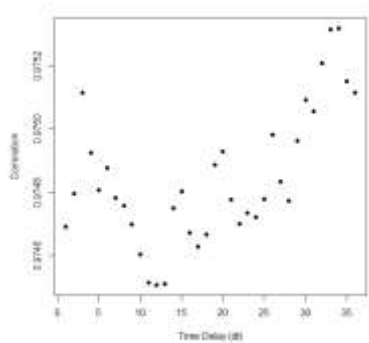

**Figure 9:** Correlation between the observed and estimated wind speed based on the buoy conditions over the give time delay.

As illustrated in Figure 8, there are oscillations in the correlation between the buoy conditions and the storm conditions. This is hard to illustrate in the relationship as there are six terms: storm **wind speed**, **storm pressure**, **buoy wind speed**, **buoy pressure**, **atmospheric** and **water temperature** (as measured at the buoy). Therefore, we will indicate the storm wind speed by size of the point character and the storm pressure by the coloring as depicted in Figure 10.

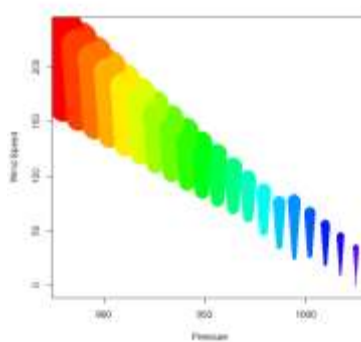

**Figure 10:** Size and Color code for Wind Speed and Pressure

Illustrated in Figure 11 (a), the wind speed in a tropical storm does not appear to be dependent on the wind speed and pressure at the buoy as indicated by the varying colors and size; however, there is a pressure below which there is a slight change in the expected storm winds. The interesting variable is water temperature (which shows oscillation) and pressure.

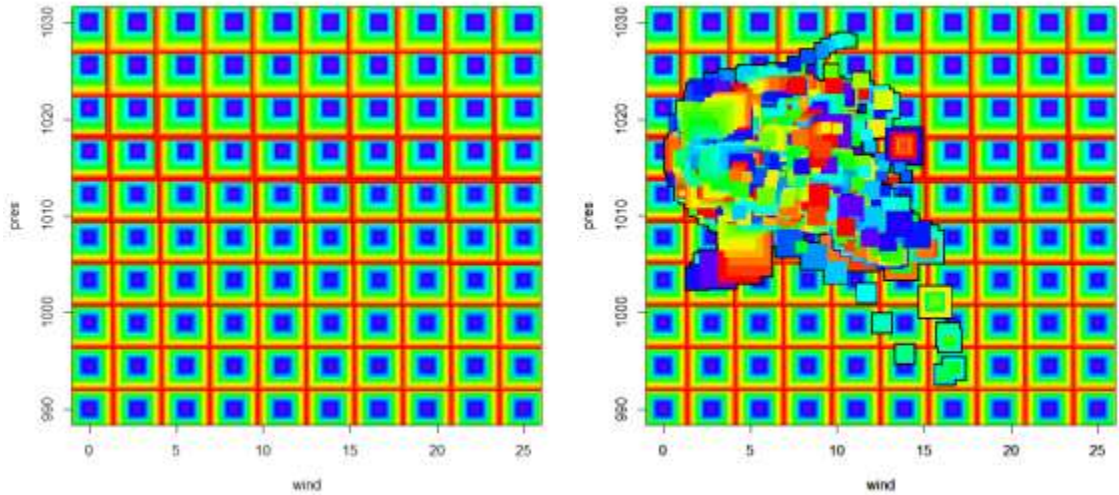

Fig. 11 (a)    Contour plot of pressure and wind speed as measured at the buoy.

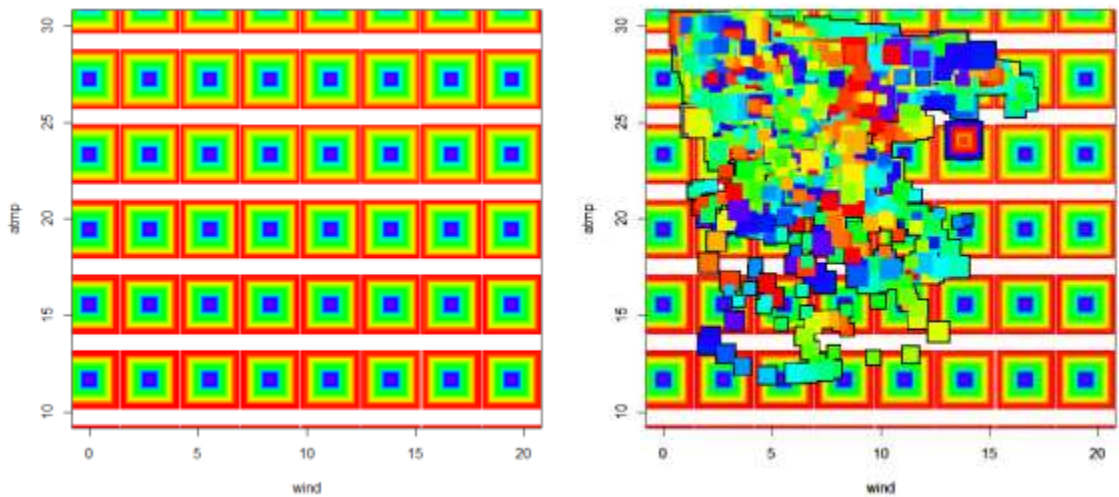

Fig. 11 (b)    Contour plot of atmospheric temperature and wind speed as measured at the buoy.

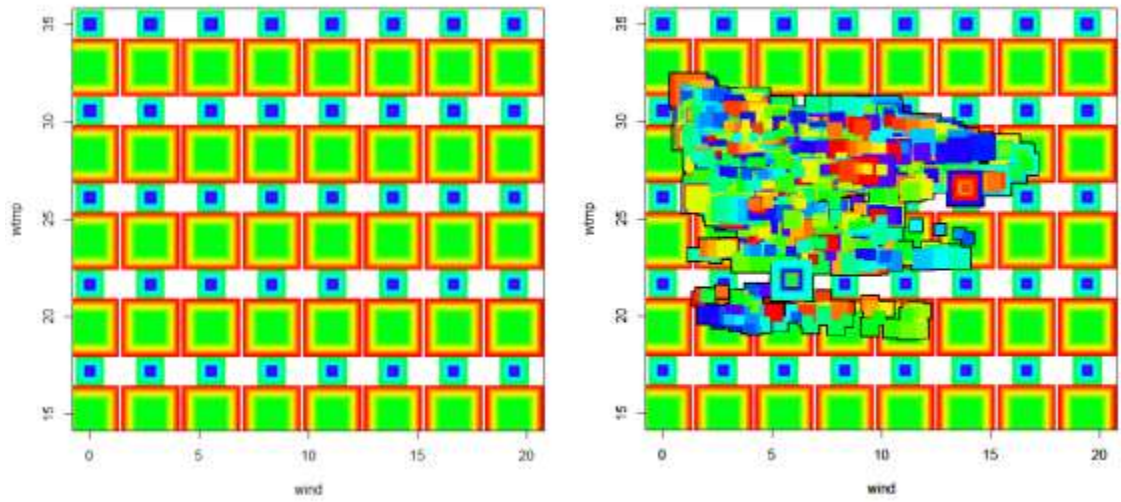

Fig. 11 (c)   Contour plot of water temperature and wind speed as measured at the buoy.

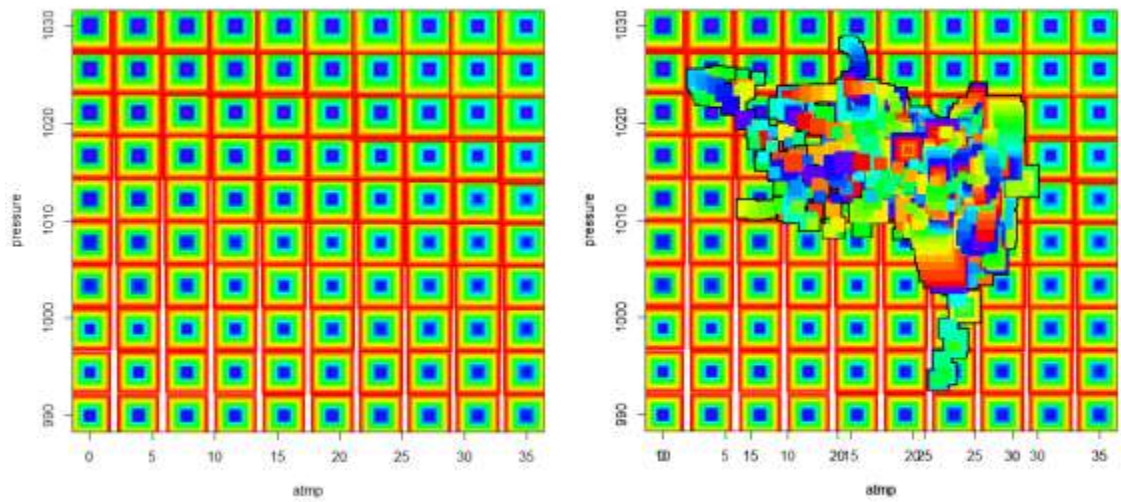

Fig. 11 (d)   Contour plot of pressure and atmospheric temperature as measured at the buoy.

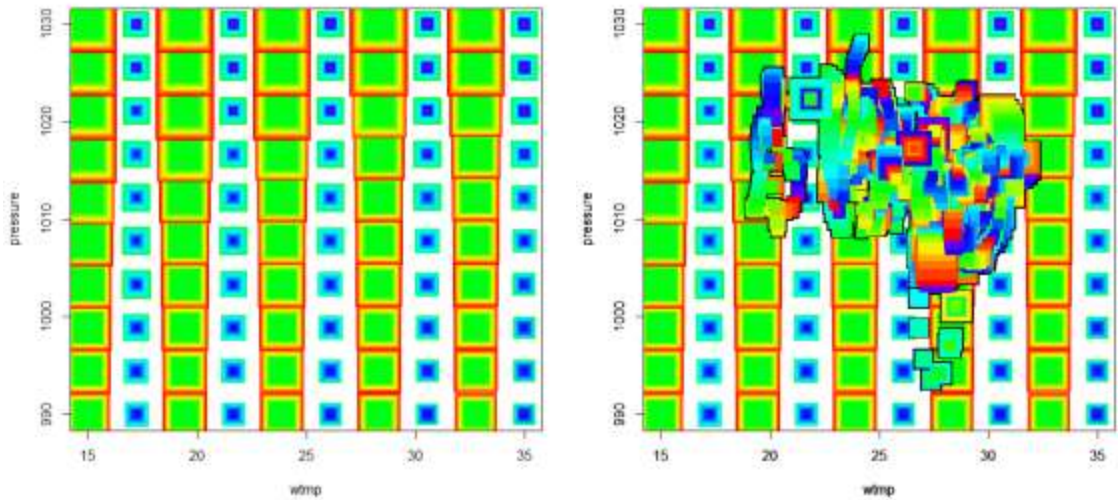

Fig. 11 (e)   Contour plot of pressure and water temperature as measured at the buoy.

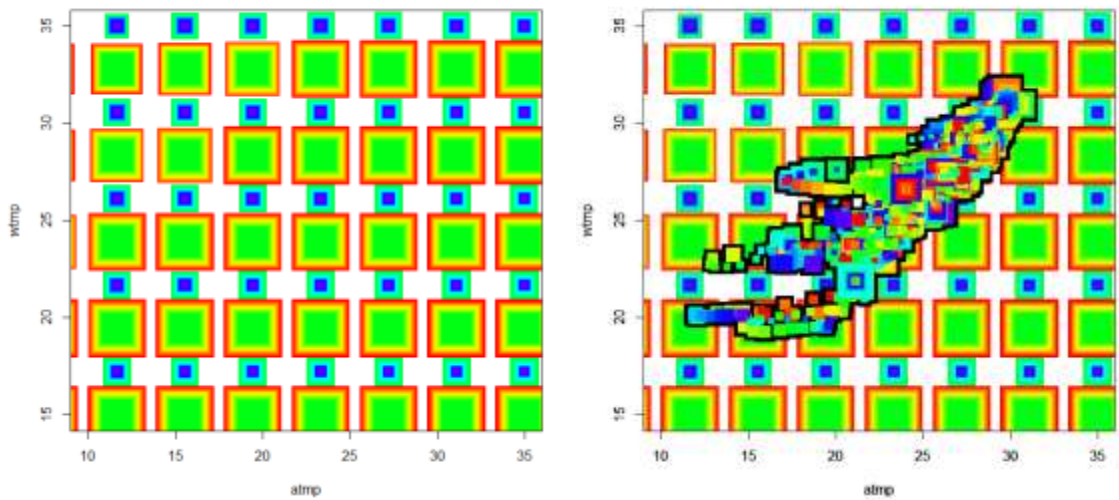

Fig. 11 (f)   Contour plot of atmospheric temperature and water temperature as measured at the buoy.

**Figure 11:**   Contour plots using size and coloring of meshed space to indicate storm winds and storm pressures in relationship to (a) buoy wind speed by buoy pressure; (b) buoy wind speed by atmospheric temperature, (c) buoy wind speed by water temperature; (d) buoy pressure by atmospheric temperature, (e) buoy pressure by water temperature; and (f) atmospheric temperature by water temperature.

This analysis indicates that there is a pumping action between pressure and wind speed as well as the non-linear relationship between storm wind speed and conditions measured at the buoy.

The unique wind speeds generated by a tropical storm and hurricanes are multiples of 5 from 10 to 160 knots, $W$. Consider the average buoy conditions by hurricane force winds (from formation through intensification and dissipation),

$$\bar{w}_i, \bar{p}_i, \bar{a}_i, \bar{t}_i, i \in W.$$

As illustrated in Figure 12, there are sinusoidal relationship in the relationship between storm winds and the conditions recorded at the buoy.

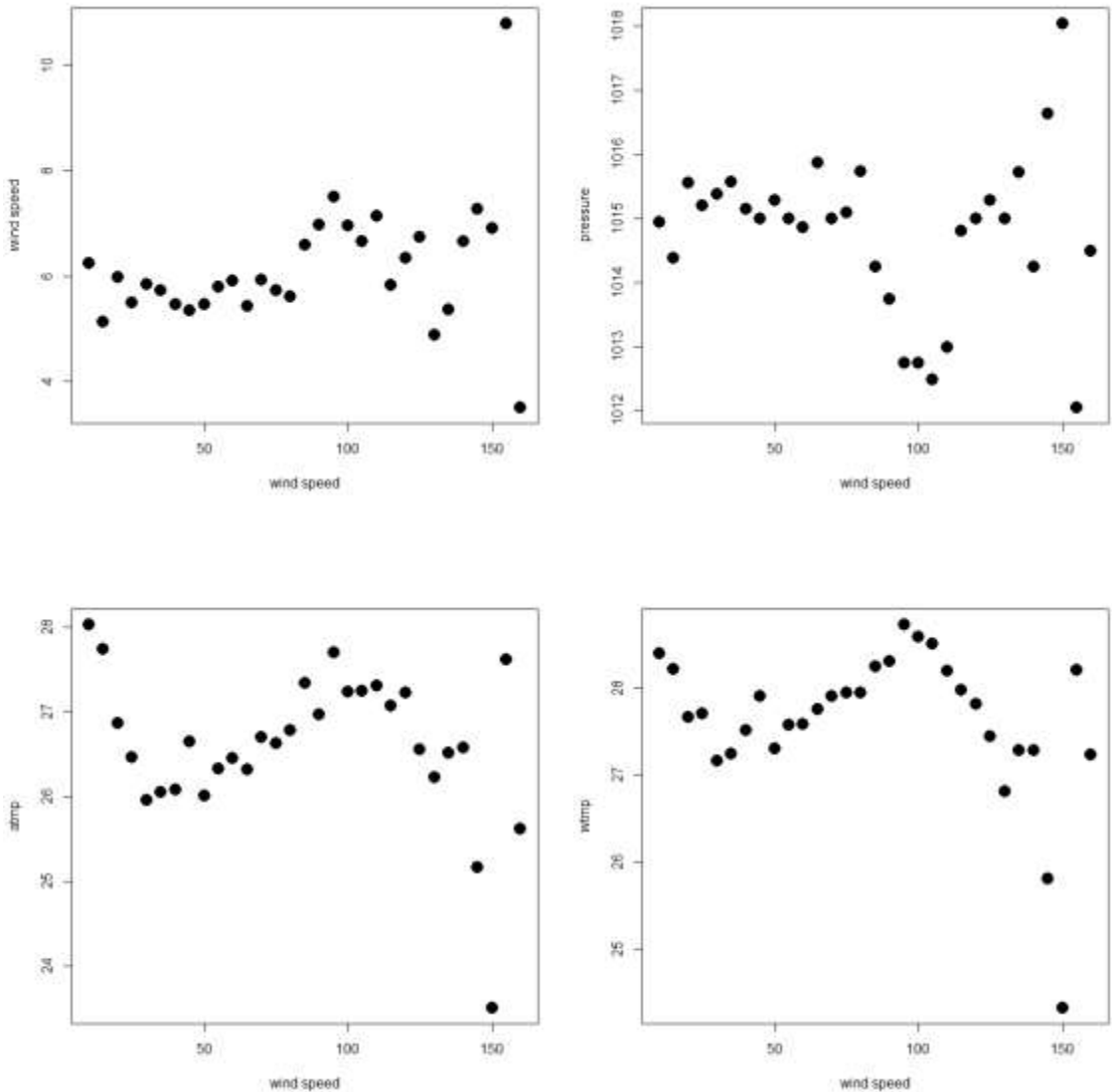

**Figure 11:** Scatter plots of storm winds and (a) wind speed at the buoy; (b) pressure at the buoy; (c) atmospheric temperature at the buoy; and (d) water temperature at the buoy.

To further investigate the apparent drop in hurricane force winds consider the buoy conditions: wind speed ($w$), pressure ($p$), atmospheric temperature ($a$) and water temperature ($t$). With coefficient of determination 0.99984, pressure is rather constant only dropping and showing higher variability when the wind speeds are high; with a coefficient of determination of 0.8624187, wind speed is the most variant. Atmospheric temperature ranks second with $R^2 = 0.9853786$ and water temperature ranks third with $R^2 = 0.92936$. What was unexpected was the clear quadratic relationship that appears between wind speed and temperature

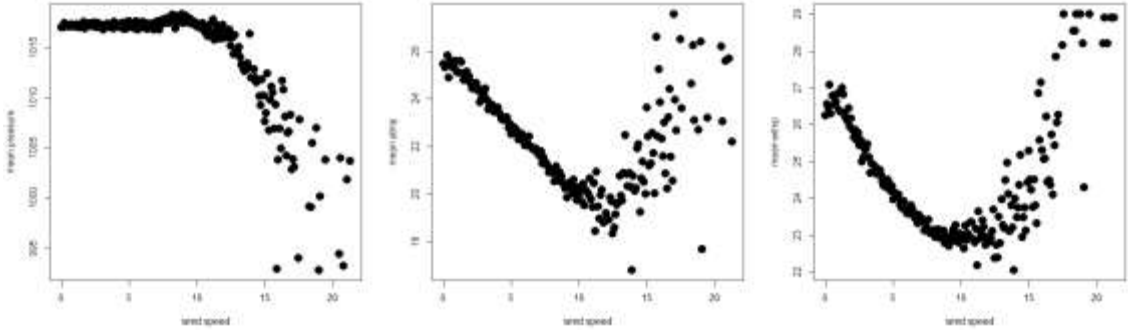

**Figure 12:** Scatter plot of buoy wind speed by (a) mean pressure, (b) mean atmospheric temperature and (c) mean water temperature.

In the non-response analysis model:

$$u = \alpha_1 p + \alpha_2 a + \alpha_3 t + \alpha_4 p^2 + \alpha_5 a^2 + \alpha_6 t^2,$$

where $u$ is a column vector of one.

Using the developed model given by

$$\hat{u} = 0.001993p - 0.00007051a + 0.0003194t - 0.000000988p^2$$
$$+ 0.0000004513a^2 + 0.000007737t^2$$

we have, as illustrated in Figure 13, that the relationship that exists is a conical, shown in black.

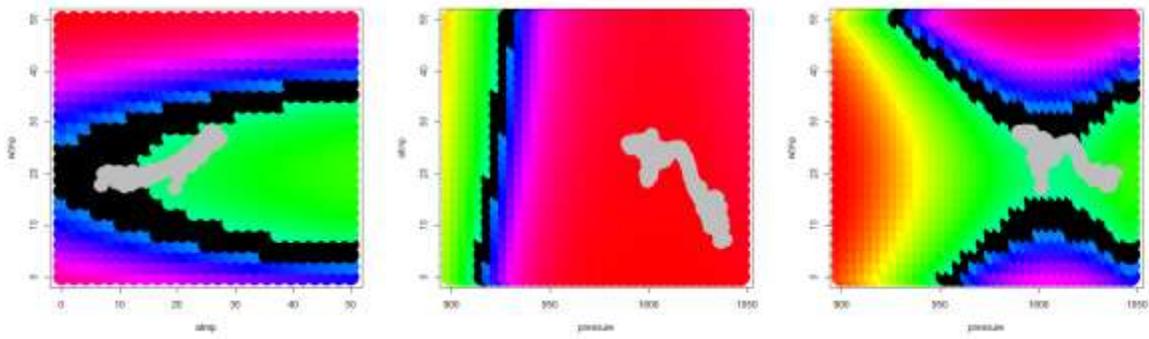

**Figure 13:** Contour plots of conic sections by estimated unity for (a) atmospheric temperature by water temperature, (b) pressure by atmospheric temperature and (c) presssure by water temperature. In the set of images, the observed data is shown in gray.

In the non-response analysis model:

$$u = \alpha_1 w + \alpha_2 p + \alpha_3 a + \alpha_4 t + \alpha_5 w^2 + \alpha_6 p^2 + \alpha_7 a^2 + \alpha_8 t^2 + \alpha_9 wp + \alpha_{10} wa + \alpha_{11} wt + \alpha_{12} pa + \alpha_{13} pt + \alpha_{14} at$$

The developed model is given by

$$\hat{u} = 0.0008058w + 0.001943p + 0.001761a - 0.0008463t + 0.00000004416w^2 \\ - 0.0000009442p^2 - 0.0000007349a^2 - 0.00000005396t^2 \\ - 0.0000007829wp - 0.000000652wa + 0.0000002316wt \\ - 0.000001708pa + 0.0000008201pt + 0.0000005655at.$$

Illustrated in Figure 14 and Figure 15, the relationship that exist is a conical shown in black.

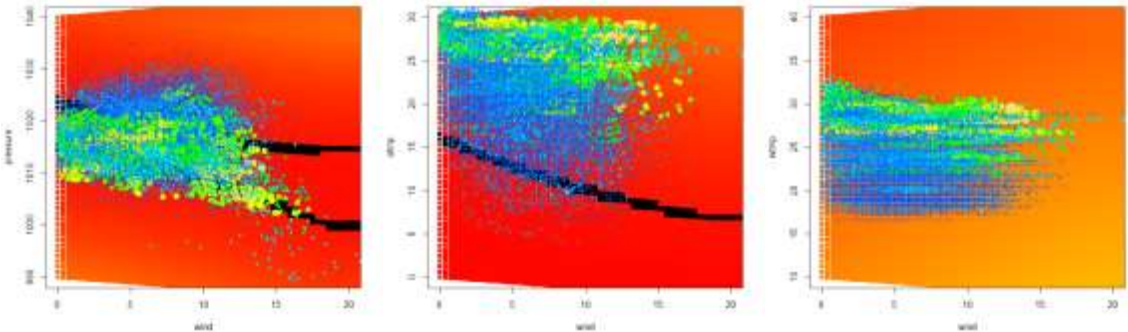

**Figure 14:** Contour plots of conic sections by estimated value unity for (a) wind speed and pressure, (b) wind speed and atmospheric temperature and (c) wind speed and water temperature.

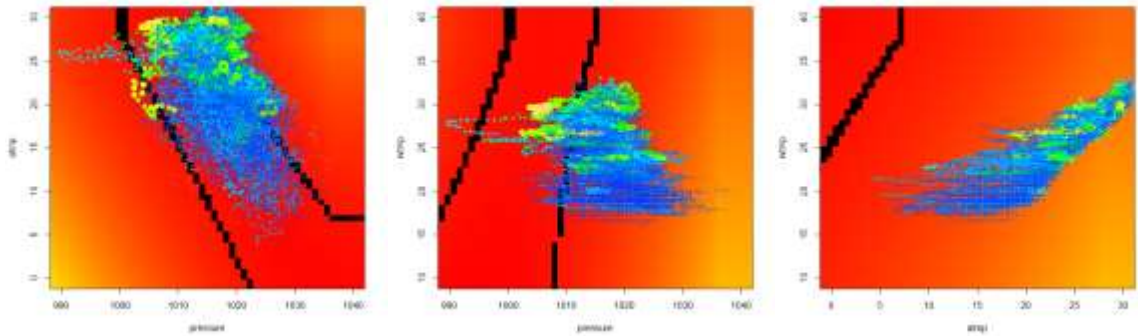

**Figure 15:** Contour plots of conic sections by estimated value of unity for (a) atmospheric temperature and pressure, (b) water temperature and atmospheric temperature and (c) atmospheric temperature and water temperature.

## 4. Usefulness

This analysis is useful in the field of meteorology as it allows co-dependent relationships among atmospheric conditions to be expressed implicitly. It also illustrates the constant ebb and flow of each of these measures in an effort to maintain a stable system. The analysis also shows that the formation of a storm may be detected days before a storm forms. The end result of this analysis will be an application which reads the current conditions at the buoy and predict the formation of a tropical storm based on the conditions near the surface of the water.